\documentclass[journal=apchd5,manuscript=letter]{achemso}

\usepackage{hyperref}
\usepackage{graphicx}
\usepackage{xspace}
\usepackage{color}
\usepackage{siunitx}
\usepackage{xcolor}
\usepackage[left, modulo]{lineno}
\usepackage{blindtext}
\usepackage{acronym}
\newcommand{\siv}{\ac{siv}\xspace}
\newcommand{\sn}{Si$_3$N$_4$}
\newcommand{\nd}{\ac{nd} }

\author{Niklas~Lettner}
\affiliation{Institute for Quantum Optics, Ulm University, 89081 Ulm, Germany}
\alsoaffiliation{Center for Integrated Quantum Science and Technology (IQST), Ulm University, Albert-Einstein-Allee 11, 89081 Ulm, Germany}
\altaffiliation{These authors contributed equally.}
\author{Lukas~Antoniuk}
\affiliation{Institute for Quantum Optics, Ulm University, 89081 Ulm, Germany}
\altaffiliation{These authors contributed equally.}

\author{Anna~P.~Ovvyan}
\affiliation{Institute of Physics and Center for Nanotechnology, University of M\"unster, D-48149 M\"unster, Germany}
\alsoaffiliation{Kirchhoff-Institute for Physics, Heidelberg University, Im Neuenheimer Feld 227, 69120 Heidelberg, Germany}
\author{Helge~Gehring}
\affiliation{Institute of Physics and Center for Nanotechnology, University of M\"unster, D-48149 M\"unster, Germany}
\author{Daniel~Wendland}
\affiliation{Institute of Physics and Center for Nanotechnology, University of M\"unster, D-48149 M\"unster, Germany}
\alsoaffiliation{Kirchhoff-Institute for Physics, Heidelberg University, Im Neuenheimer Feld 227, 69120 Heidelberg, Germany}
\author{Viatcheslav~N.~Agafonov}
\affiliation{GREMAN, UMR 7347 CNRS, INSA-CVL, Tours University, 37200 Tours, France}
\author{Wolfram~H.~P.~Pernice}
\affiliation{Kirchhoff-Institute for Physics, Heidelberg University, Im Neuenheimer Feld 227, 69120 Heidelberg, Germany}
\alsoaffiliation{Institute of Physics and Center for Nanotechnology, University of M\"unster, D-48149 M\"unster, Germany}
\alsoaffiliation{SoN — Center for Soft Nanoscience, 48149, M\"unster, Germany}
\author{Alexander~Kubanek}
\email{Alexander.Kubanek@uni-ulm.de}
\affiliation{Institute for Quantum Optics, Ulm University, 89081 Ulm, Germany}
\alsoaffiliation{Center for Integrated Quantum Science and Technology (IQST), Ulm University, Albert-Einstein-Allee 11, 89081 Ulm, Germany}

\title{Controlling all Degrees of Freedom of the Optical Coupling in Hybrid Quantum Photonics}
\abbreviations{IR,NMR,UV}

\begin{document}
\newacro{qpp}[QPP]{quantum post-processing}
\newacro{pl}[PL]{photoluminescence}
\newacro{nd}[ND]{nanodiamond}
\newacroplural{nd}[NDs]{nanodiamond}
\newacro{pcc}[PCC]{photonic crystal cavity}
\newacro{pccs}[PCCs]{photonic crystal cavities}
\newacroplural{pcc}[PCCs]{photonic crystal cavity}
\newacro{afm}[AFM]{atomic force microscope}
\newacro{siv}[SiV$^-$-center]{negatively-charged silicon-vacancy center }
\newacro{opo}[OPO]{optical parametric oscillator}
\newacro{fwhm}[FWHM]{full width at half maximum}
\newacro{psb}[PSB]{phonon sideband}
\newacro{ple}[PLE]{photoluminescence excitation}
\newacro{ldos}[LDOS]{local density of states}
\newacro{zpl}[ZPL]{zero-phonon line}
\newacro{na}[NA]{numerical aperture}
\newacro{fdtd}[FDTD]{Finite Difference Time Domain}

\begin{tocentry}
\includegraphics[scale=1]{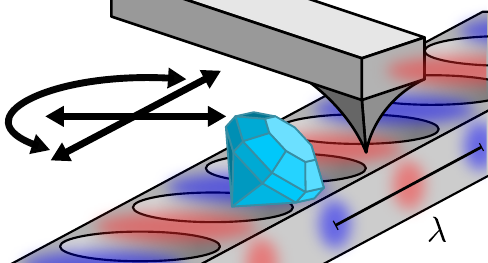}

Title: Controlling all Degrees of Freedom of the Optical Coupling in Hybrid Quantum Photonics

Authors: Niklas Lettner, Lukas Antoniuk, Anna P. Ovvyan, Helge Gehring, Daniel Wendland, Viatcheslav N. Agafonov, Wolfram H. P. Pernice, Alexander Kubanek

Brief Synopsis: The graphic shows the AFM based nanomanipulation technique utilized to optimize the coupling strength by controlling the position and rotation of the nanodiamond containing the silicon vacancy center relative to the photonic crystal cavity.

\end{tocentry}

\begin{abstract}
Nanophotonic quantum devices can significantly boost light-matter interaction which is important for applications such as quantum networks. Reaching a high interaction strength between an optical transition of a spin system and a single mode of light is an essential step which demands precise control over all degrees of freedom of the optical coupling. While current devices have reached a high accuracy of emitter positioning, the placement process remains overall statistically, reducing the device fabrication yield. Furthermore, not all degrees of freedom of the optical coupling can be controlled limiting the device performance. Here, we develop a hybrid approach based on negatively-charged silicon-vacancy center in nanodiamonds coupled to a mode of a \sn - photonic crystal cavity, where all terms of the coupling strength can be controlled individually. We use the frequency of coherent Rabi-oscillations and line-broadening as a measure of the device performance. This allows for iterative optimization of the position and the rotation of the dipole with respect to individual, preselected modes of light. Therefore, our work marks an important step for optimization of hybrid quantum photonics and enables to align device simulations with real device performance.
\end{abstract}

\section{Introduction}

The development of efficient spin-photon interfaces is an ongoing research challenge required to establish quantum technologies such as quantum networks \cite{rufQuantumNetworksBased2021}. The essential step is to achieve best-possible optical coupling between a spin system and a photonic channel. The coupling strength to individual modes can be drastically increased when the emitter is placed in the field maximum of an optical resonator mode, this enhancement is known as the Purcell effect \cite{zhangStronglyCavityEnhancedSpontaneous2018}. Among the large variety of optical resonators, \ac{pccs} integrated on-chip, are especially promising for a scalable technology due to a small footprint, high-throughput fabrication and good performance due to high quality factors and small mode volumes \cite{elshaariHybridIntegratedQuantum2020a,ovvyanElectroluminescentTunableCavityenhanced2023}. Independent of the photonic platform, the optimization of the optical coupling strength still remains an ongoing challenge. \\
Hybrid quantum photonics opens new avenues for further optimization of spin-photon interfaces with the potential to control and optimize the coupling strength. Hybrid quantum photonics with color centers in a \ac{nd} \cite{sahooHybridQuantumNanophotonic2023} enables to separately optimize the spin system, here a \ac{siv} in a \ac{nd}, and the photonics platform, namely \ac{pccs} in \sn-photonics. In a final step, referred to as \ac{qpp}  \cite{kubanekHybridQuantumNanophotonics2022}, both entities are connected by means of optical coupling. Importantly, high-throughput integration processes are being developed to enable large-scale hybrid quantum photonic devices with simultaneous, multi-channel access \cite{schrinnerIntegrationDiamondBasedQuantum2020} and heterogeneous integration based on, for example, standard CMOS foundry processes \cite{wengHeterogeneousIntegrationSolid2023}. Further improvements arise from interaction zones, which are introduced in the photonics material prior to the \ac{qpp} \cite{frochPhotonicNanobeamCavities2020, alagappanPurcellEnhancementLight2020}. Nevertheless, \ac{qpp} relies on high-precision yet typically statistical processes which does not enable to optimize the coupling strength. 

In this work, we demonstrate the capability to control all degrees of freedom in the \ac{qpp} step in order to enable, in principle, deterministic coupling of an optical transition of the \siv to individual modes of light defined by the electric field of an integrated \ac{pcc}. Therefore, we utilize pre-selected \siv in a \ac{nd} and place it at characteristic positions within the simulated electric field distribution of a \sn-based \ac{pcc}. \sn-\siv hybrid quantum photonics has been developed over the past years demonstrating efficient integration with strong Purcell-enhancement \cite{fehlerPurcellenhancedEmissionIndividual2020, fehlerHybridQuantumPhotonics2021} and recently with access to the \siv electron spin $via$ a Purcell-broadened optical transition \cite{antoniukAllOpticalSpinInitialization2023}. Here, we focus on the \ac{qpp} and demonstrate the deterministic nature of the AFM-based nanomanipulation \cite{hausslerPreparingSingleSiV2019}.

\section{Results} 

The quality of a quantum light-matter interface is often quantified by comparing its performance with regard to the free-space properties of the quantum emitter. Considering the free-space emission, the corresponding emission rate is given by the spontaneous decay rate $\Gamma$. The emission enhancement of an emitter coupled to a single mode of a cavity is given by the formula

\begin{equation}\label{eq:1}
	\frac{\Gamma_{\mathrm{cav}}}{\Gamma} = F_{\mathrm{cav}}\cdot \frac{1}{1+4Q^2\left(\frac{\lambda_{\mathrm{SiV}}}{\lambda_{\mathrm{cav}}}-1\right)^2  }\cdot\left(\frac{\vec{E} (\vec{r})\cdot\vec{\mu}}{|\vec{E}_ {\mathrm{max}}||\vec{\mu}|}\right)^2 
\end{equation}
and can be separated into three different terms \cite{englundControllingSpontaneousEmission2005}. The first being the Purcell factor $F_{\mathrm{cav}}$. The second term accounts for the spectral overlap between the cavity mode and the optical transition of the emitter. The third term encodes the spatial overlap between the dipole and the electric field of the cavity mode, as well as the dipole orientation. In order to maximize the coupling, simultaneous control over all three coupling terms is required. Our hybrid approach allows us to separately control all three terms. Each term will be discussed individually in the following paragraphs.

\subsection{Purcell-factors of the \ac{pcc} modes}
\begin{figure}[htpb]
	\centering
	\includegraphics[scale=1]{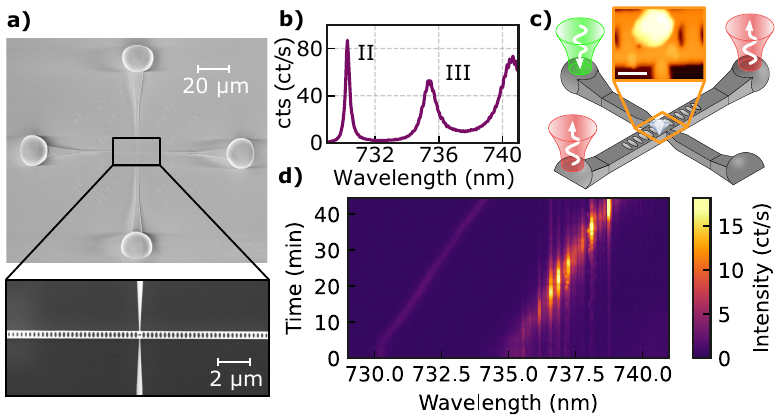}
	\caption{Assembled hybrid quantum system \textbf{a)} SEM image of the \sn -photonic circuit with a cross-bar \ac{pcc} design to suppress the background fluorescence  for off-resonant excitation \cite{EfficientCouplingEnsemble}, 1D \ac{pcc} formed by two non-uniform Bragg mirrors and a cavity region in between. Broadband 3D printed couplers are used to couple light in and out of the waveguides.  \textbf{b)} The measured \ac{pcc} transmission spectrum of the II-mode at \SI{730}{\nano\meter} and the III-mode at \SI{735}{\nano\meter}. \textbf{c)} The schematic of an ND containing \siv placed on the cross-bar PCC device. The system is excited off-resonantly with a \SI{532}{\nano\meter} laser $via$ the pump beam. The \ac{pl} emission of the coupled system can be detected out of the cavity couplers. \textbf{d)} The measured \ac{pl} signal, $\Gamma_{\mathrm{cav}}$, monitored over time during the gas tuning process shows distinct \siv transitions coupled to the III-mode.}
	\label{fig:coupling_ensemble}
\end{figure}

The Purcell factor, $F_{\mathrm{cav}} = \frac{3}{4\pi^2}\frac{\lambda^3}{n^3}\frac{Q}{V}$, is determined by the cavity parameters and can be controlled and maximized by optimization of PCC design and high-quality, precise fabrication. Our photonic platform is based on a cross-bar design with a 1D \ac{pcc} formed by two identical non-uniform Bragg mirrors with a cavity region in between, as shown in Fig. \ref{fig:coupling_ensemble} a). The cross-bar design allows for off-resonant excitation $via$ the pump beam while suppressing the background florescence of the \sn \space in the cavity beam\cite{fehlerEfficientCouplingEnsemble2019}. A on-substrate \ac{pcc}-design in comparison to freestanding structures enables further post-processing $via$ \ac{afm} access and decreases fabrication complexity. The \ac{pcc} device is equipped with broadband, total-reflection base, three-dimensional (3D) couplers connected to the ends of the nanophotonic waveguides to excite the emitter and read-out the optical signal, see SI 1 for more details \cite{gehringBroadbandOutofplaneCoupling2019}. The device used in this work exhibits six distinct cavity modes. We utilize the II- and III-mode, which both extend further into the cavity beam in contrast with the I-mode, to demonstrate both rough and fine position adjustments of the ND relative to the PCC. The transmission spectrum of the II- and III-mode is shown in Fig. \ref{fig:coupling_ensemble} b). The experimental $Q$-factors of $Q_{\mathrm{II}}=2200$ for the II-mode and $Q_{\mathrm{III}}=800$ for the III-mode in combination with the simulated mode volumes $V_{\mathrm{II}}=5.8\frac{\lambda^3}{n^3}$ and  $V_{\mathrm{III}}=6.2\frac{\lambda^3}{n^3}$ are used to calculate the theoretically achievable Purcell factors of $F_{\mathrm{cav,II}} = 29$ and $F_{\mathrm{cav,III}} = 10$, respectively.

\subsection{Controlling the spectral overlap}

The second term of the coupling equation (Eq. \ref{eq:1}) accounts for the spectral overlap between the cavity resonance frequency and the optical transition frequency of the quantum emitter. We transfer a \ac{nd} (lateral size $\approx$ \SI{350}{\nano\meter}), which contains an ensemble of SiV$^-$-centers, to the \ac{pcc} by means of \ac{afm}-based pick and place technique \cite{fehlerHybridQuantumPhotonics2021,schellScanningProbebasedPickandplace2011}. For details on the ND synthesis see SI 2. The sample is placed inside a helium flow cryostat and cooled to $\approx$ \SI{4}{\kelvin}. The emitter is excited in a home build confocal microscope (SI 3) through the crossed pump beam, while the cavity coupled emission is detected at the right 3D coupler. (see. Fig. \ref{fig:coupling_ensemble} c). Next, we record the emission spectra while tuning the cavity resonances with the help of a controlled nitrogen gas flow \cite{mosorScanningPhotonicCrystal2005}. The observation of different transitions coupling to the III-mode, as shown in Fig. \ref{fig:coupling_ensemble} d), visualize the effect of the detuning on the coupling strength. In this example, the frequencies of the cavity resonances shifts linearly by an amount of \SI{0.85}{\giga\hertz\per\second}. The tuning speed can be controlled by means of a ultra-high vacuum gas dosing valve and can be chosen significantly lower, if needed. When the resonance condition is fulfilled, the positioning of the cavity resonances can be stabilized within \SI{10}{\giga\hertz}, limited by the resolution of the spectrometer which is used to determine the detuning. For finer tuning precision continuous laser scanning could be utilized. We achieve a long-term stability within the spectrometer resolution over 8 hours. The stability of the resonance frequency together with the fine control over the detuning, results in a control over the detuning term to within \SI{99}{\percent} possible coupling. This is calculated by substituting the obtained values into the second term of Eq. \ref{eq:1}.

\subsection{Field overlap and dipole alignment}
\begin{figure}[htpb]
	\centering
	\includegraphics[scale=0.95]{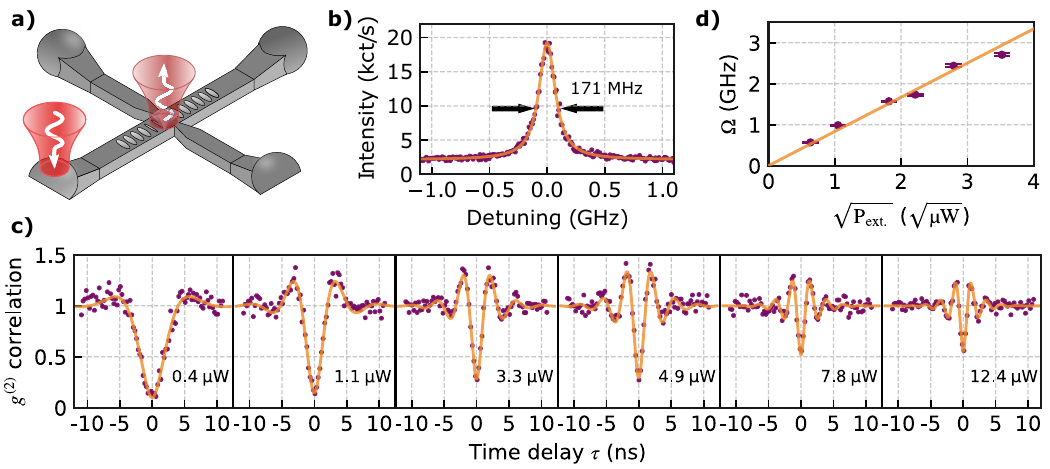}
	\caption{Measurement of the coherent optical driving of the coupled system by means of second-order correlations $g^{(2)}(\tau)$. \textbf{a)} The excitation laser is coupled into the cavity mode $via$ the 3D coupler while the \ac{psb} is detected in free-space from the position of the \ac{nd}. \textbf{b)} \ac{ple} scan of the emitter with a central emission wavelength of \SI{736.05}{\nano\meter} excited with a power of \SI{500}{\nano\watt}. \textbf{c)} Power dependent correlation measurements $g^{(2)}(\tau)$. \textbf{d)} Corresponding linear scaling of the Rabi-frequency in relation to the square root of the excitation power.}
	\label{fig:resonant_excitation}
\end{figure}

To quantify the third coupling term of Eq. \ref{eq:1}, which accounts for the dipole alignment and electric field strength at the emitter position, it is important to isolate a single emitter exhibiting a promising coupling strength in the \ac{pl} measurement. Resonant excitation enables to probe the coupling strength of an individual transition. The \ac{ple} measurements were performed by coupling the laser into the cavity mode, thereby increasing the field strength at the emitter position, as compared to free-space excitation, $via$ cavity enhancement (Fig. \ref{fig:resonant_excitation} a). The \ac{psb} emission is detected at the location of the \ac{nd}. The spectral line at \SI{736.05}{\nano\meter} exhibits a high count rate and a narrow linewidth, shown in Fig. \ref{fig:resonant_excitation} b). 

\begin{figure}[htpb]
	\centering
	\includegraphics[scale=1]{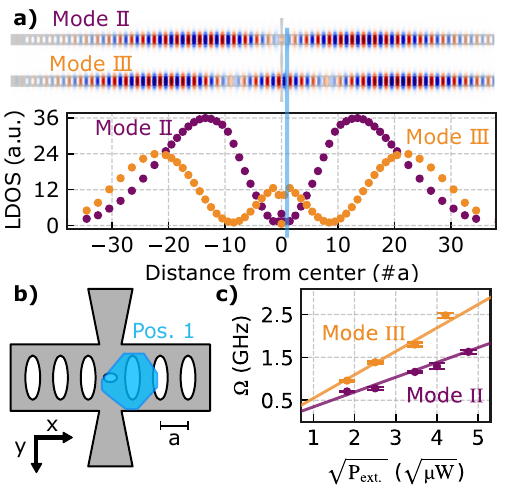}
	\caption{Coarse positioning of the emitter on the \ac{pcc} \textbf{a)} Simulated electric field distribution $ E_y $ for the II- and III-mode along the cavity beam. The spatially simulated \ac{ldos} is plotted below against the distance from the center in units of $ a $ (distance between holes). \textbf{b)} Sketch of the \ac{nd} placed at the center of the PCC. \textbf{c)} Scaling of the Rabi-frequency for the ND placed at Pos. 1 coupled to the II- and III-mode.} 
	\label{fig:rough_positioning}
\end{figure}

\begin{figure}[htpb]
	\centering
	\includegraphics[scale=1]{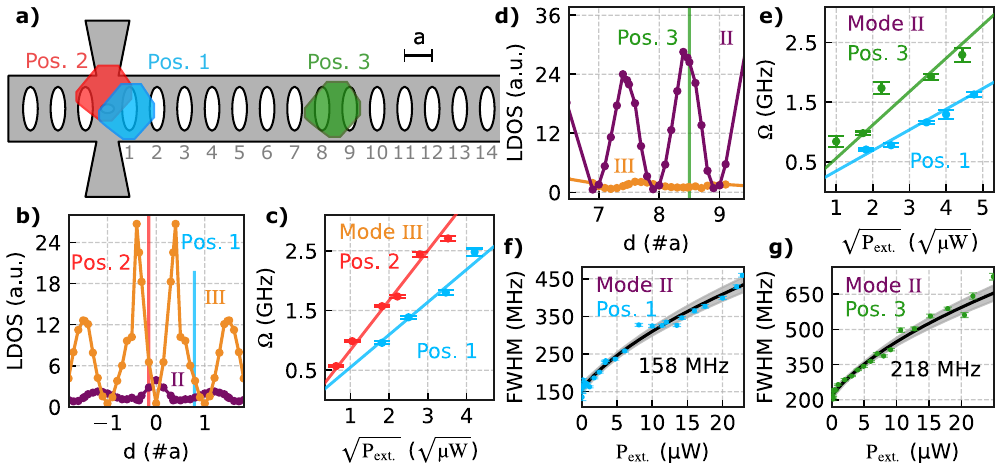}
	\caption{Optimization of the coupling strength. \textbf{a)} Sketch of the three \ac{nd} positions which are studied to demonstrate the optimization of the coupling strength. \textbf{b)} \ac{ldos} simulated for the II- and III-mode in the center of the device plotted against the distance d from the center of the cavity in units of the hole spacing a.  \textbf{c)} Rabi-scaling for the III-mode at Pos. 1 and Pos. 2. \textbf{d)} \ac{ldos} simulated for the II- and III-mode from hole 7 to 9. \textbf{e)} Rabi-scaling for the coupling to the II-mode at Pos. 1 and Pos. 3. \textbf{f)} Power-dependent linewidth measurement in resonance with the II-mode at Pos. 1, results in $ \delta_0=\SI{158\pm5}{\mega\hertz}. $ \textbf{g)} Power-dependent linewidth measurement at Pos. 3 in resonance with the II-mode yielding $ \delta_0=\SI{218\pm6}{\mega\hertz} $.}
	\label{fig:fine_positioning}
\end{figure}

The single quantum emitter nature is verified by second-order correlation measurements, $g^{(2)}(\tau)$, resulting in $g^{(2)}(0)=$\SI{0.10\pm0.03}< 0.5 (Fig. \ref{fig:resonant_excitation} c) with \SI{400}{\nano\watt} excitation power. When a two level system is strongly driven coherent oscillations between the excited and ground state can be observed, called Rabi-oscillations. The Rabi-frequency is proportional to $\vec{E} (\vec{r})\cdot\vec{\mu}$, directly allowing conclusions to be drawn on the third coupling term (Eq. \ref{eq:1}). 
The Rabi-frequency can be extracted from the $g^{(2)}(\tau)$-measurement by fitting 

\begin{equation}\label{key}
	g^{(2)}(\tau) = 1 - \left(\frac{SNR}{SNR+1}\right)^2 e^{-3/4\Gamma|\tau|}\left(\cos(\Omega_d|\tau|)+\frac{3\Gamma}{4\Omega_d}\sin(\Omega_d|\tau|)\right) 	
\end{equation}
with $\Omega_d = \sqrt{\Omega^2-\left(\frac{\Gamma}{4}\right)^2}$, where $\Omega$ is the Rabi-frequency, $SNR$ is the signal to noise ratio and $\Gamma$ is the spontaneous decay rate of the system \cite{steckQuantumAtomOptics2007}. We measure excitation-power dependent second-order correlations, where the Rabi-frequency scales proportional to the square root of the excitation power. For an excitation power of \SI{12.4}{\micro\watt}, measured before the objective, we obtained a Rabi-frequency of $\Omega=\SI{2.7}{\giga\hertz}$. Fig. \ref{fig:resonant_excitation} d) shows the measured Rabi-frequencies against the square root of excitation power with a slope \SI{0.82\pm0.03}{\frac{\giga\hertz}{\sqrt{\micro\watt}}}, when the optical transition is coupled to the III-mode. While the Rabi-frequency for a fixed excitation power could be used as a measure of $\vec{E} (\vec{r})\cdot\vec{\mu}$ we further increase the accuracy by using the scaling of the Rabi-frequency.
By measuring the Rabi-frequency scaling for different emitter positions and orientations within the cavity field, it is possible to determine the product $\vec{E} (\vec{r})\cdot\vec{\mu}$ while keeping $|\vec{E}_ {\mathrm{max}}||\vec{\mu}|$ constant. We use \ac{fdtd} to simulate the electric field distribution and \ac{ldos} of the respective modes to determine the position dependent coupling strength (Fig. \ref{fig:rough_positioning} a). 
The simulations of the $E_y$-field distribution of the cavity clearly reveals the standing wave of the cavity field modulated by an envelope. For the II-mode the electric field is symmetric around the center of the cavity with a node in the center of the device. The III-mode is symmetric around the center with two nodes along the cavity. In a first step, we perform rough positioning of the \ac{nd} in the center of the device. For the first position (Pos. 1), sketched in Fig. \ref{fig:rough_positioning} b), we determined the Rabi-scaling for the II- and III-mode. This results in a slope of \SI{0.33\pm0.01}{\frac{\giga\hertz}{\sqrt{\micro\watt}}} for the II-mode and a slope of \SI{0.56\pm0.02}{\frac{\giga\hertz}{\sqrt{\micro\watt}}} for the III-mode. This is in agreement with the simulated \ac{ldos} (Fig. \ref{fig:fine_positioning} b) which is significantly lower for II-mode compared to the III-mode at this position. With the help of fine positioning of the ND the coupling to both modes can be improved.
For the III-mode we therefore adjust the position of the \ac{nd} on a scale smaller than size of the \ac{nd}, within the anti-node of the envelope (Fig. \ref{fig:rough_positioning} a), to optimize the position within the standing wave. The \ac{nd} is positioned at the center of the device as sketched in Fig. \ref{fig:fine_positioning} a) (Pos. 2). The \ac{ldos} simulations in Fig. \ref{fig:fine_positioning} b) hint to an increase in coupling strength for coupling to the III-mode. This is verified by comparing the respective Rabi-scalings. The optimization results in an increase of the Rabi-scaling from \SI{0.56\pm0.02}{\frac{\giga\hertz}{\sqrt{\micro\watt}}} to \SI{0.82\pm0.03}{\frac{\giga\hertz}{\sqrt{\micro\watt}}} at Pos. 2 (Fig. \ref{fig:fine_positioning} c). The II-mode exhibits a higher Purcell enhancement factor $F_{\mathrm{cav}}$, making it the more promising mode to achieve high coupling strength.
To increase the coupling of the emitter to the II-mode we move the \nd from the node of the mode envelope (Fig. \ref{fig:rough_positioning} a) closer to the anti-node by moving the \ac{nd} to inbetween the $8^{th}$ and $9^{th}$ cavity hole. For Pos. 3 we expect to see no coupling to the III-mode due to the envelope node, however the \ac{ldos} for the II-mode is significantly higher in comparison to Pos. 1 (Fig. \ref{fig:fine_positioning} d). This optimization step yields a significant increase in Rabi-scaling of \SI{0.56\pm0.05}{\frac{\giga\hertz}{\sqrt{\micro\watt}}} compared to \SI{0.33\pm0.01}{\frac{\giga\hertz}{\sqrt{\micro\watt}}} at Pos. 1  (Fig. \ref{fig:fine_positioning} e).
All Rabi-scalings are measured in relation to the excitation power $\mathrm{P_{ext}}$ in front of the objective. Since the transmission of the II-mode is lower compared to the III-mode it is difficult to reliably compare the Rabi-scalings between the II- and III-mode.
We can verify the increase in coupling strength by measuring the line-broadening of the emitter. Fitting the power-dependent linewidths with $\delta = \delta_0\sqrt{1+\mathrm{P_{ext}}/\mathrm{P_{sat}}}$ enables to extract the linewidth at zero power $\delta_0$. For Pos. 1 we obtain $ \delta_{0,\mathrm{Pos. 1}}=\SI{158\pm5}{\mega\hertz} $ in resonance with the II-mode (Fig. \ref{fig:fine_positioning} f). The optimization of the emitter to Pos. 3 results in a line-broadening of $ \Delta_{\mathrm{1 \rightarrow 3}}=\SI{60\pm8}{\mega\hertz} $ to $ \delta_{0,\mathrm{Pos. 3}}=\SI{218\pm6}{\mega\hertz} $ (Fig. \ref{fig:fine_positioning} g). This increase in linewidth with respect to the free-space linewidth can be used to determine the cooperativity. To determine the free-space linewidth, we rotate the \ac{nd} orthogonal to the cavity axis, turning off the cavity-emitter coupling, yielding $ \delta_{0,\mathrm{off}}=\SI{142\pm 4}{\mega\hertz} $. We verified negligible emitter-cavity coupling by moving the \ac{nd} off the cavity beam (SI 4) resulting in $ \delta_{0,\mathrm{off beam}}=\SI{146\pm16}{\mega\hertz}$. The cooperativity of our system is determined by 
\begin{equation}\label{key}
	C = \frac{4g^2}{\kappa\gamma} = \frac{\Gamma}{\gamma}
\end{equation}
where $g$ is the single-photon Rabi-frequency, $\kappa$ the cavity decay rate,  $\gamma$ the spontaneous emission rate and $\Gamma = \delta_{0,\mathrm{on}}-\delta_{0,\mathrm{off}}$.
At Pos. 1 this results in a cooperativity of $C_{\mathrm{Pos.1}}=\SI{0.12\pm0.04}{}$, which is increased by more than a factor of four to $C_{\mathrm{Pos.3}}=\SI{0.54\pm 0.05}{}$ at Pos. 3. 

In addition to the position optimization steps we further optimize the coupling of the emitter by adjusting the rotational degree of freedom. 
\begin{figure}[htpb]
	\centering
	\includegraphics[scale=1]{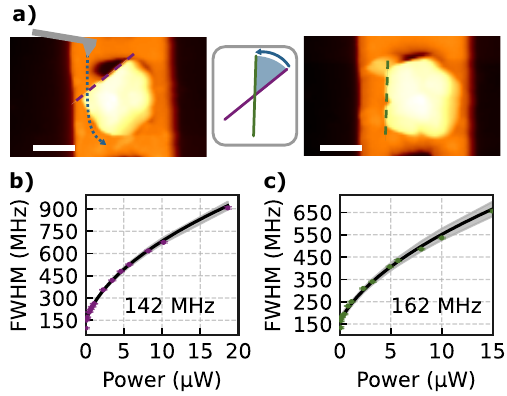}
	\caption{Optimization of the dipole alignment \textbf{a)} AFM images showing the rotation step resulting in a rotation of the \ac{nd} by \SI{49}{\degree}. The scale bar represents \SI{200}{\nano\meter}. \textbf{b)} Power-dependent linewidth of the coupled spectral line before the rotation of the ND resulting in a fitted zero-power linewidth of $ \delta_0=\SI{142\pm5}{\mega\hertz} $. \textbf{c)} Power-dependent linewidth of the coupled spectral line after rotation optimization yields a fitted zero-power linewidth of $ \delta_0=\SI{162\pm5}{\mega\hertz} $.} 
	\label{fig:rotation}
\end{figure}
Fig. \ref{fig:rotation} a) shows the first rotation step of the ND with a rotation of \SI{49}{\degree}. After this first step the ND is further rotated by \SI{32}{\degree}. The total rotation results in an increase of the linewith from $ \delta_0=\SI{142\pm5}{\mega\hertz} $ before the rotation (Fig. \ref{fig:rotation} b) to $ \delta_0=\SI{162\pm5}{\mega\hertz} $ (Fig. \ref{fig:rotation} c) coupled to the II-mode. 
Notably, the overall $Q$-factor dropped from $ Q\approx2200$ down to $Q\approx860$ due to damage on the cavity between the measurements, see SI 5 for details. The combination of fine position optimization and rotation can be used in order to itteratively optimize the emitter-cavity coupling. However, this process currently involves removing the chip from the cryostat for each individual step, which is time-consuming. While this process can be used to increase the coupling strength, as discussed above, we now propose a strategy to minimize the number of necessary steps.

\begin{figure}[htpb]
	\centering
	\includegraphics[scale=1]{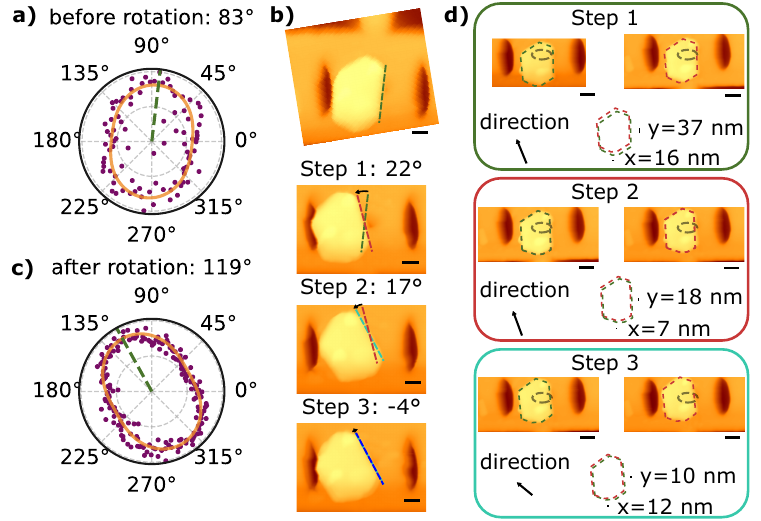}
	\caption{Rotation of the \ac{siv}-containing \ac{nd} by means of \ac{afm} -based nanomanipulation \textbf{a)} The polarization measurement before the rotation shows polarization angle of \SI{83}{\degree}. \textbf{b)} Three step nanomanipulation procedure resulting in respective changes of the \ac{nd} orientation by \SI{22}{\degree} followed by \SI{17}{\degree} and \SI{-4}{\degree}. The total rotation derived from the \ac{afm} scans is \SI{35}{\degree}. \textbf{c)} The polarization measurement after the nanomanipulation procedure yields an dipole angle of \SI{119}{\degree} in agreement with the rotation of the \ac{nd}. \textbf{d)} Three nanomanipulation steps and the relative movements in x- and y-direction in respect to the center of the cavity. The scalebar represents \SI{100}{\nano\meter}.}
	\label{fig:rotation_afm}
\end{figure}
Therefore, we first adjust the dipole alignment relative to the cavity by measuring the polarization of the emitter (Fig. \ref{fig:rotation_afm} a) resulting in a angle of \SI{83}{\degree}. Please note, that we utilize a different \ac{nd} here, with a characteristic crystal shape to demonstrate the procedure. We then rotate the \ac{nd} in three steps resulting in respective changes of the \ac{nd} axis of \SI{22}{\degree} followed by \SI{17}{\degree} and \SI{-4}{\degree} (see Fig. \ref{fig:rotation_afm} b). This adds up to a total rotation of the \ac{nd} relative to the cavity mode by \SI{35}{\degree}. Measuring the polarization after the nanomanipulation steps, shown in Fig. \ref{fig:rotation_afm} c), confirms the corresponding change of the dipole axis of the quantum emitter by \SI{36}{\degree}. This technique give us the ability to align the in-plane dipole axis of the quantum emitter with the cavity axis with an estimated precision of \SI{5}{\degree}. After the dipole is aligned with the cavity axis, the emitter must be placed in a position with high field strength. This can be done with fine position optimization steps as shown in Fig. \ref{fig:rotation_afm} d), where the position changes between steps are as small as \SI{10}{\nano\meter} without changing the orientation of the ND. Finally, the electric field at the emitter position could iteratively be determined from the Rabi-oscillations to optimize the coupling strength with a position accuracy of \SI{10}{\nano\meter}. In combination with the simultaneous control of cavity resonances frequency by gas tuning and the device design, this enables the control of the coupling strength between a pre-characterized quantum emitter and the cavity mode with a minimal number of optimization steps.

\section{Conclusion}

In this manuscript, we demonstrate experimental control over all degrees of freedom of the optical coupling term for our hybrid quantum photonics platform. The hybrid quantum system is assembled by placing an \siv-containing \nd onto a \sn-photonic device with an \ac{afm}-based pick and place technique. Isolating a single optical transition out of an \siv ensemble $via$ resonant excitation, enables probing of the local electric field strength at the emitter position. Therefore, cavity enhanced resonant excitation yields fast coherent driving of a single \siv. We utilize the scaling of the resulting coherent Rabi-oscillations as a measure of the local electric field strength. In combination with \ac{afm}-based nanomanipulation we map the field distribution of the cavity modes by continuously monitoring the coherent Rabi-oscillations. This allows us to optimize the coupling strength to a specific resonance. We demonstrate access to both position and rotation of the dipole with respect to the field maximum of the cavity mode. Rotations of the \ac{nd} relative to the cavity mode can be used both to suppress or enhance the coupling strength. By measuring the polarization of a single emitter relative to the cavity mode polarization, it is possible to align the in-plane dipole with an estimated precision of \SI{5}{\degree} corresponding to \SI{99}{\percent} ($\cos(5^{\circ})$) of the maximal coupling strength. Iteratively measuring the coupling strength between each optimization step allows us to position the emitter relative to the cavity field with an accuracy much better than the size of the \ac{nd} host.  When the dipole is aligned to the cavity it is possible to move the \ac{nd} relative to the cavity with an accuracy of \SI{10}{\nano\meter}. Optimization of the position yields an increase of the Rabi-scaling in resonance with the II-mode by a factor of 1.6. The increase in coupling strength is further verified by comparing the spectral linewidths, while mitigating power-broadening effects. For the II-mode, we see a linewidth increase of $ \Delta_{\mathrm{1 \rightarrow 3}}=\SI{60\pm7}{} $ to $ \delta_{0,\mathrm{Pos. 3}}=\SI{218\pm6}{\mega\hertz} $. The coupling term corresponding to the detuning between the optical transition frequency of the \siv and the resonance frequency of the cavity mode is controlled to within \SI{99}{\percent} of the maximum possible coupling strength. Finally, we determine the cooperativity of the coupled system by rotating the dipole orthogonal to the cavity axis resulting in complete decoupling of the emitter from the cavity mode. We achieved an optimized cooperativity of $C_{\mathrm{Pos.3}}=\SI{0.54\pm 0.05}{}$.

The developed \ac{qpp} enables the deterministic coupling of pre-selected quantum emitters to pre-selected photonics. The \siv is especially interesting for the use in small host crystals due to its inversion symmetry \cite{xuFabricationSingleColor2023}. Expanding this tool-chain to smaller \ac{nd} and to intrinsic coupling within the mode field maximum \cite{fehlerHybridQuantumPhotonics2021} in combination with higher $Q$-factors will enable cooperativities well above 1. Recent developments of spin access $
$ Purcell-broadened optical transitions \cite{antoniukAllOpticalSpinInitialization2023} and shown prolonged orbital relaxation \cite{klotzProlongedOrbitalRelaxation2022} open the door for spin-dependent device reflectivity in the strong coupling regime \cite{bhaskarExperimentalDemonstrationMemoryenhanced2020}. Furthermore, recently shown two-photon interference between two remote NDs \cite{waltrichTwophotonInterferenceSiliconvacancy2023a} paves the way for distributing quantum information between remote quantum nodes \cite{bersinDevelopmentBostonarea50km2023,pompiliRealizationMultinodeQuantum2021}. While the optimization procedure is not yet a high-throughput operation, it nevertheless marks a crucial step for final device optimization and could be scaled-up by automatization. Furthermore, the demonstrated technique enables to adjust  numerically simulations to real operation parameters of the final system. For example, the simulated field distribution is prone to error arising from fabrication imperfections which could now be mapped out on the composite system.

\section{Acknowledgments}
The work done in this manuscript is funded by the BMBF/VDI in the project HybridQToken.  A.K. acknowledges support of the Baden-Wuerttemberg Stiftung gGmbH in Project No. BWST-ISF2018-008. A.K. acknowledges support of the BMBF/VDI in the project Spinning.
We thank Prof. Kay Gottschalk and Carolin Grandy for the support and possibility to utilize the AFM, funded by the DFG. N.L. acknowledges support of the IQST. H.G. acknowledges the Studienstiftung des deutschen Volkes for financial support. The authors thank V.A. Davydov for synthesis and processing of the nanodiamond material. We thank P. Maier for production of the marker structures. D.W. acknowledges funding from the Deutsche Forschungsgemeinschaft (CRC 1459). Experiments are controlled with the open-source software Qudi \cite{binderQudiModularPython2017}.

\bibliography{Driving}

\end{document}